%% file: main_revised.tex
\documentclass[final,5p,times,numbers]{elsarticle}

\usepackage{amssymb}
\usepackage{amsmath}
\usepackage{comment}

\usepackage{tabularx}
\usepackage{booktabs}
\usepackage{hyperref}
\usepackage{siunitx}   
\usepackage[colorinlistoftodos, color=green!40, prependcaption]{todonotes}
\usepackage{scrextend} 
\usepackage{multirow}
\usepackage{svg}
\newcommand{\half}{\frac{1}{2}}

\journal{arXiv}

\graphicspath{{figures}}

\begin{document}

\begin{frontmatter}

\title{High resolution large working distance scanning helium microscopy}

\author[cam,isis]{SM Lambrick\corref{cor1}} 
\author[cam]{NA von Jeinsen}
\author[cam]{A Radi\'{c}}
\author[cam,ion]{DJ Ward}
\author[glas]{JE Halpin}
\author[glas]{DA MacLaren}
\author[cam]{AP Jardine}
\cortext[cor1]{Corresponding author: \url{sam.lambrick@stfc.ac.uk}}

\affiliation[cam]{organization={Cavendish Laboratory, University of Cambridge},
            addressline={JJ Thompson Avenue}, 
            city={Cambridge},
            postcode={CB3 0US}, 
            country={UK}}
\affiliation[ion]{organization={Ionoptika Ltd.},
            addressline={Unit B6, Millbrook Close}, 
            city={Chandler’s Ford},
            postcode={SO53 4BZ}, 
            country={UK}}
\affiliation[glas]{organization={SUPA, School of Physics \& Astronomy, University of Glasgow},
			city={Glasgow},
			postcode={G12 8QQ},
			country={UK}}
\affiliation[isis]{organization={ISIS Facility, Rutherford Appleton Laboratory, Chilton},
			city={Didcot},
            state={Oxfordshire},
			postcode={OX11 0QX},
			country={UK}}

\begin{abstract}
Scanning helium microscopy (SHeM) is attractive for imaging delicate and insulating surfaces because it combines a non-destructive neutral-atom probe with strong surface sensitivity. Recent pinhole microscopy experiments highlight that practical constraints relating to the equipment dimensions often limit resolution. Notably, both the intensity and the effective spot size of a divergent He beam are improved by a short working distance; but short working distances are restricted by the physical size of the vacuum system required to support, and detect a helium beam. Here, we report sub-micron resolution whilst maintaining a large-working-distance, with an intrinsic beamwidth of \SI{340}{\nano\metre} achieved at working distances of \SI{770}{\micro\metre} to \SI{850}{\micro\metre}. This sixfold improvement over our previous configuration is enabled by a new sample chamber with a movable source, which allows optimisation of the atom optics, together with a more compact pinhole mounted directly beside a larger detector aperture. Geometric contributions to the measured beamwidth from the pinhole are equal to those from the demagnified source size, and slightly larger than diffraction contributions, placing the instrument in a near-optimised regime. The resulting combination of sub-micron beam size, useful depth of field and practical sample access is demonstrated on bacterial specimens and eroded diamond. The work establishes large-working-distance pinhole SHeM as a viable sub-micron imaging platform and extends its usefulness for topographic imaging and micro-diffraction applications.
\end{abstract}


\end{frontmatter}

\section{Introduction}

High-resolution scanned beam surface microscopy is not defined solely by the ability to form a small probe. In practice, usable resolving power depends on the ability to recover contrast-bearing information from neighbouring regions of a specimen with sufficient signal relative to noise and background. However, achieving usable high resolution is particularly challenging for delicate, insulating, or topographically complex samples, where charged-particle techniques may suffer from beam damage, charging, or the need for conductive coatings\cite{egerton_radiation_2004,postek_does_2013}, while optical methods may lack the required surface specificity,  spatial resolution, or depth of field. Scanning helium microscopy (SHeM) addresses part of this gap by using a neutral helium atom beam as a non-destructive surface probe\cite{palau_neutral_2023,eder_sub-resolution_2023,von_jeinsen_2d_2023}.

SHeM forms images by rastering a sample beneath a narrow neutral helium beam and detecting the scattered flux. Because helium atoms in these instruments typically have incident energies of only $10$--$\SI{70}{\milli\electronvolt}$, they interact with the outer electronic density of the surface rather than penetrating into the material or driving the damage pathways more commonly associated with charged beams\cite{holst_material_2021}. At the same time, their de Broglie wavelength, typically $0.5$--$\SI{2}{\angstrom}$, remains on the atomic scale, giving access to diffraction-based sensitivity to very small structural features\cite{von_jeinsen_2d_2023,hatchwell_measuring_2024}. In practice, therefore, while the ultimate physical limit is set by the probe wavelength, the usable resolution is set by the practical atom optics and signal constraints of the instrument. A number of SHeM instruments have been developed and, while based on similar principles, these can be broken down by their method of generating the narrow helium beam: using an active focusing zone plate\cite{flatabo_reflection_2024}, or using pinhole collimation. Pinhole instruments can then be further broken down into two classes: those that use a large working distance and skimmed beam, such as the instrument used in the current work\cite{barr_design_2014,bhardwaj_neutral-atom-scattering-based_2022} or those using a short working distance and un-skimmed beam\cite{witham_simple_2011}. The short-working-distance instruments have, to date, demonstrated the highest resolution, beamwidths $315-\SI{350}{\nano\metre}$, while the more flexible long-working-distance instruments have remained restricted to beamwidths $>\SI{1}{\micro\metre}$. Addressing this gap is the principle aim of the current work.

The unique characteristics of helium as a microscopy probe have led to growing interest in SHeM for classes of specimen and measurement that benefit from strong surface sensitivity combined with exceptionally low probe energy. Recent work has demonstrated the utility of helium microscopy for 2D materials~\cite{bhardwaj_neutral-atom-scattering-based_2022,bhardwaj_contrast_2023,radic_measuring_2025}, biological structures~\cite{myles_taxonomy_2019,lambrick_multiple_2020}, and other samples prone to charging or beam-induced modification. Beyond topographic imaging, the same instrumental platform can also be adapted for microscopic atom-diffraction measurements\cite{von_jeinsen_2d_2023,hatchwell_measuring_2024}. The combination of non-destructive imaging, surface specificity and diffraction sensitivity is one of the distinctive attractions of helium microscopy.

Usable resolution reflects the combined effect of: 1. source brightness, 2. detection efficiency, and 3. the imaging or beam-forming optics that deliver signal to a defined region of the sample\cite{bergin_standardizing_2022}. The first two elements are common to all microscopies, and advances in both beam generation\cite{eder_focusing_2012,eder_velocity_2018} and efficient detection\cite{bergin_low-energy_2021} have been important in making high-performance helium microscopy experimentally viable. In particular, the high resolution configuration presented in the current work relies on a step change in detector efficiency in the form of a custom high sensitive helium detector developed by Bergin et al.\cite{bergin_low-energy_2021} which has an efficiency $\gtrsim\times 100$ greater than commercially available detectors. However, the present work focuses on the third element: the atom-optical geometry of a large-working-distance pinhole SHeM, and the trade-offs between beam size, beam intensity, depth of field, and practical sample access. Therefore we are concerned not only with reducing beamwidth -- though beamwidth and edge-response measurements remain a key quantifiable performance metric -- but with improving the experimentally useful resolving capability of large-working-distance SHeM.

The resolution gap between large-working-distance pinhole SHeM and the best reported beamwidth is critical because the practical advantages of larger working distances -- including tolerance of significant sample topography, greater depth of field, and instrument modularity -- are many of the features that make SHeM attractive for broader microscopy use. To address the resolution gap, we apply previously developed theoretical atom-optical optimisations\cite{bergin_method_2019,salvador_palau_theoretical_2017} to the Cambridge SHeM, showing that the trade-off between working distance and beamwidth can be shifted substantially. We report sub-micron resolution in a large-working-distance pinhole SHeM, achieving an intrinsic beamwidth of $\SI{340}{\nano\metre}$ while maintaining working distances greater than $\SI{700}{\micro\metre}$. This corresponds to a six-fold improvement over our previous large-working-distance configuration. The advance is enabled by combining constrained optimisation of the atom optics with a redesigned sample chamber, improved pinhole fabrication, a reduced pinhole diameter, an increased source-pinhole distance, and a compact high-resolution `pinhole-plate'  geometry (the pinhole-plate being the defining optical element in long working distance SHeM\cite{bergin_complex_2021}). In addition, the ability to maintain usable signal levels at larger working distances is also aided by the higher source brightness of a skimmer-collimated beam and a high efficiency solenoidal detector not originally present on the Cambridge SHeM\cite{bergin_low-energy_2021}. Further, we demonstrate the improved operating regime  using examples from biological imaging and nanostructured materials. Thus, we establish large-working-distance pinhole SHeM as a practical sub-micron imaging platform and a more capable basis for future topographic and micro-diffraction measurements.

\section{Large-working-distance SHeM}

Several implementations of neutral-beam real-space microscopy have now been demonstrated, but they differ substantially in both atom-optical design and practical operating regime. These include the Bergen neutral atom microscope (NAM), which uses Fresnel zone-plate focusing\cite{koch_imaging_2008,flatabo_reflection_2024,eder_focusing_2012,eder_focusing_2015,flatabo_atom_2017}, the NAM developed by Witham et al.\ at Portland State University\cite{witham_simple_2011,witham_increased_2012}, and the Cambridge\cite{barr_design_2014,zhao_multi-detector_2025}, Newcastle\cite{fahy_highly_2015} and Hyderabad SHeMs\cite{bhardwaj_neutral-atom-scattering-based_2022} that follow the pinhole-collimation principles introduced by Barr et al.\cite{barr_design_2014,zhao_multi-detector_2025}. Because the optical constraints and performance trade-offs are markedly different for zone-plate and pinhole instruments, the present work concerns only pinhole-collimated systems\cite{bergin_method_2019,salvador_palau_theoretical_2017}.

\begin{figure}
    \centering
    \includegraphics[width=\linewidth]{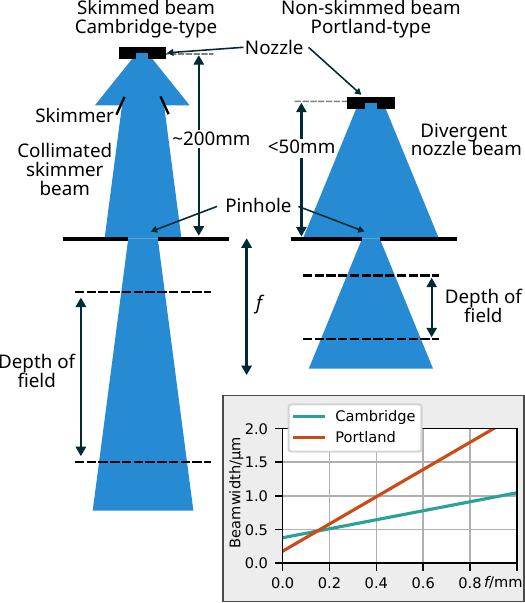}
    \caption{Basic geometry of the skimmed beam long working distance Cambridge-type SHeM and the non-skimmed beam short working distance Portland-type SHeM. The Cambridge-type uses a collimated skimmer beam as its source, providing a small divergence after pinhole collimation. The Portland-type directly uses the nozzle beam close to the pinhole, providing a large flux with a much more divergent beam. A more divergent beam will result in a smaller depth of field as the beamwidth will rapidly grow too large, as shown in the inset graph (the inset is demonstrative and does not represent any specific instrument configurations).}
    \label{fig:shem_v_nam}
\end{figure}

The terminology in the literature is not always fully consistent: both \emph{SHeM} and \emph{NAM} have been used either as generic labels for helium atom imaging or as names for specific instruments. In addition the term NAM has been applied to a broader category of microscopy where atomic species other than helium may been employed\cite{witham_exploring_2014,bhardwaj_contrast_2023,zhao_multi-detector_2025} and, in principle, could include non-scanning approaches to image formation. Here, we use \emph{SHeM} as the general term for scanning beam microscopy with a neutral helium beam and refer to specific instruments by their host institution. The two pinhole-based instrument classes, represented in figure \ref{fig:shem_v_nam}, will be referred to as the non-skimmed beam short working distance \emph{Portland-type} and skimmed beam long working distance \emph{Cambridge-type} -- we note these type descriptors do not comment on the capability of either the Portland NAM or Cambridge SHeM instruments to be adapted to operate in different configurations, but represent broad design categories based on the published configurations. A key distinction between the two types lies in the  geometry, particularly that of the incident optics. The Portland-type uses a non-skimmed source beam, allowing a small distance between the nozzle and pinhole, and typically operating at short working distances ($30-\SI{100}{\micro\metre}$\cite{witham_increased_2012}) in order to minimise the beamwidth and maximise signal,  but sacrificing depth of field and flexibility of operating conditions. Meanwhile the Cambridge-type operates with a much larger source to pinhole distance and with a skimmer, subsequently allowing operation at substantially larger working distances ($>\SI{1}{\milli\metre}$ previously and $>\SI{600}{\micro\metre}$ in the current work), providing a significantly larger depth of field, and flexibility for other types of measurement, such as spot profile diffraction\cite{von_jeinsen_2d_2023} and 3D profilometry\cite{radic_heliometric_2025}.

The extended working distance, $f$, used by the Cambridge-type SHeM confers several benefits, the most significant being: a much larger depth of field and tolerance for larger 3D topography. We note that unlike with active focusing optics, pinhole optics have no sharp focal plane around which a depth of field is naturally defined, instead an \emph{effective depth of field} arises from the spreading of the beam from the pinhole and the detection geometry, and is centred on the coincident point of the incident beam and the detection cone. Above a certain $f$ the beam will have spread out too much to be useful, while the detection geometry limits the range of $f$ over which a usable signal can be acquired\cite{lambrick_ray_2018}, in our previous work for the Cambridge-type SHeM we found the effective depth of field to be $\pm\frac{1}{3}f_D$ where $f_D$ is the design working distance/coincident point\cite{lambrick_ray_2018}. We note that such a coincident point is difficult to define in the Portland-type geometry.
Further, the added space the larger working distance allows has enabled the Cambridge-type to be a modular instrument, with replaceable defining optical elements, termed `pinhole-plates', modifying the instrument for different incidence angles\cite{radic_3d_2024}, angular resolutions\cite{von_jeinsen_2d_2023} and reduced background\cite{bergin_complex_2021} with only a single component change.  In addition, the larger distance from the nozzle to the sample leads to a more monochromatic beam\cite{deponte_brightness_2006} and hence provides momentum, $k$, resolution, allowing adaptation of the instrument for atom-diffraction measurements on microscopic sample regions\cite{von_jeinsen_2d_2023,hatchwell_measuring_2024}. 
Thus, the large working distance of the Cambridge-type geometry is not merely a convenience, but one of its central experimental strengths. These advantages have previously, however, as already notes, come at the expense of spatial resolution with Cambridge-type implementations have restricted to beamwidths $>\SI{1}{\micro\metre}$,
while beamwidths $315-\SI{350}{\nano\metre}$ reported for Portland-type implementation using a working distance of $\sim\SI{30}{\micro\metre}$\cite{witham_increased_2012}. The discrepancy lies largely in the fact that earlier Cambridge-type optical configurations had not been fully optimised. The optimisation frameworks developed by Bergin et al.\ and Salvador-Palau et al.\ provide the means to address this directly\cite{bergin_method_2019,palau_theoretical_2016,salvador_palau_theoretical_2017}, and the current work reports the realisation of the atom-optic optimisation.

\section{Optical Optimisation and Trade-offs}
\label{sec:optical_optimisation}
\subsection{Beam width}

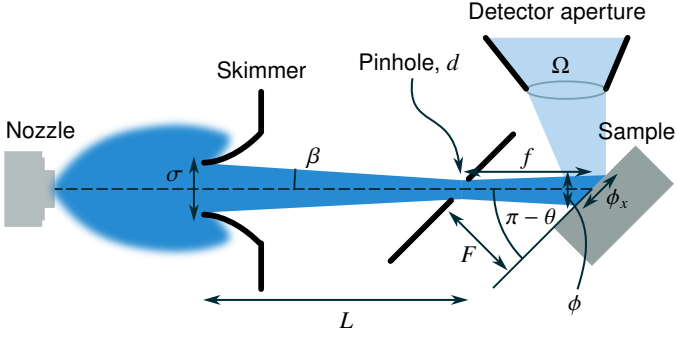
\begin{figure}
    \centering
    \textsf{\small\input{optical_system.tex}}
    \caption{The optical system in the Cambridge SHeM. High pressure gas forms a supersonic expansion after passing through a nozzle into a vacuum. The centre-line of the expansion is selected by a skimmer. The beam is then passed through a pinhole aperture and is projected onto the sample, which lies at $\ang{45}$ to the helium beam. $\phi$ is the intrinsic beamwidth (perpendicular to the beam propagation direction) while $\phi_x$ is the beamwidth projected onto the scanning plane.}
    \label{fig:optical_system}
\end{figure}

The optical system of the Cambridge SHeM is outlined schematically in figure \ref{fig:optical_system}. A supersonic expansion of helium is generated by exposing high pressure gas to vacuum via a narrow nozzle; the centreline of the expansion is selected by a skimmer producing a near monochromatic helium beam \cite{kelsall_minimizing_2025}. The beam is then passed through a pinhole, performing the final collimation and demagnification of the source. Finally, it is incident on the sample at an angle of $\theta=\ang{45}$ where the helium scattered in a particular direction is collected by a detector aperture. Key physical parameters of the system that can be changed are: the diameter of the pinhole, $d$, the working distance, $f$, the distance between the source and the pinhole, $L$, and the solid angle of the detector aperture, $\Omega$. The virtual source size (FWHM), $s$, largely determined by the skimmer diameter in the Cambridge SHeM\cite{bergin_instrumentation_2018}, is left fixed, and the helium wavelength is also a fixed parameter. The angular source size, $\beta$, is dependent on $s$ and $L$.

Bergin\cite{bergin_method_2019} and Salvador-Palau\cite{palau_theoretical_2016} identified three main contributors to the width of a helium beam at the sample in a pinhole microscope: the geometric projection of the pinhole, the demagnified source size, and diffraction through the pinhole. Both optimisation methods assume that the contributions from each of these are approximately Gaussian, giving the total beamwidth at the sample as a quadrature sum of the three contributions. 

Both theoretical methods considered a normal incidence microscope, however our instrument operates in a $\ang{45}$ incidence configuration, and therefore modifications to the mathematics are necessary. In addition, one of the key parameters of the optimisation is the angular source size, $\beta$. Unlike some other parameters, the angular source size does not directly correspond to a single physical parameter we have control over. In the current work we use a known source, with a previously measured virtual source size, and change the source--pinhole distance in order to change $\beta$. Thus, in the small angle approximation, $\beta=s/L$, where $L$ is the distance between the source and the pinhole and $s$ is the measured virtual source size.
Modifying Bergin's formula for the beam standard deviation for non-normal incidence, see \ref{sec:non-normal_incidence}, and substituting $\beta=s/L$, we arrive at 
\begin{equation}\label{eq:beamwidth_use}
    \phi_x^2 = \underbrace{\left(\frac{d}{2\sqrt{3}}\right)^2}_{\text{geometric}} +  \underbrace{\left(\frac{1}{\cos\theta}\frac{s f}{\sqrt{3}L}\right)^2}_{\text{source size}} +  \underbrace{\left(\frac{1}{\cos\theta} \frac{0.42\lambda f}{d\cos\theta}\right)^2}_{\text{diffraction}},
\end{equation}
as the beam standard deviation along the horizontal scanning direction, denoted $\phi_x$ in figure \ref{fig:optical_system}. The first term in equation \ref{eq:beamwidth_use} is the geometric project of the pinhole onto the sample, this is a straightforward relation between the pinhole diameter and resulting beamwidth. The second term represents the demagnification of the finite source size onto the sample and is a function of the angular source size and the working distance. A well optimised system will have the first two terms of similar magnitude. The final term comes from the Airy disc diffraction pattern formed from a circular aperture, as the helium wavelength is not a freely tunable parameter provided this term remains smaller than the other two we remain in a well optimised configuration. Meanwhile if the diffraction term is larger than the other two then the configuration is \emph{diffraction limited} and cannot be consider well optimised.

$\phi_x$ corresponds to the beam standard deviation that is measured by scanning the beam over a sharp edge, however, as can be seen in figure \ref{fig:optical_system} the beam is actually narrower than the measured quantity by $\phi=\phi_x\cos\theta$ due to non-normal incidence of the beam on the sample. We denote the beam standard deviation with lower case $\phi$ and the beam full width at half maximum (\textsc{fwhm}) as capital $\Phi$. The \textsc{fwhm} is taken to be our measure of beamwidth and hence representative of resolution in a similar manner to previous work\cite{witham_increased_2012,bergin_instrumentation_2018,sam_m_lambrick_formation_2021,eder_focusing_2012}. We term $\Phi$ the \emph{intrinsic beamwidth} and $\Phi_x$ the \emph{projected beamwidth}. The resolving power of the instrument will depend both on the intrinsic beamwidth, and the orientation of features relative to the beam, with those perpendicular to the beam resolved the best and those parallel not resolved at all.

\subsection{Optimisation}

The optimisation procedure seeks the atom-optical parameters that maximise signal for a chosen target beamwidth. For a pinhole microscope, the mathematical optimum in working distance, $f$, lies at $f=0$, but that limit is not useful experimentally. In the Cambridge-type SHeM, reducing $f$ too far would compromise one of the central practical advantages of the geometry: a large depth of field together and tolerance of significant topography. Our aim is therefore not to reach the formal zero-working-distance optimum, but to identify the best usable configuration for a sub-micron, large-working-distance instrument.

We choose a target intrinsic beamwidth of approximately $\Phi\sim\SI{300}{\nano\metre}$ for two reasons. First, this would make the Cambridge-type geometry competitive with the best reported beamwidth of the Portland-type. Second, previous theoretical work suggests that this length scale remains favourable for a pinhole instrument, rather than requiring active focusing such as a zone plate\cite{bergin_method_2019}.

The optimisation requires a choice of working distance, $f$, from which the other parameters follow. We impose two practical constraints. First, we require a depth of field at least $\times1000$ larger than the target beamwidth. Previous studies indicate that the effective depth of field in Cambridge-type SHeM extends approximately $\pm g$ from the design working distance/coincidence point, with $g\sim f/3$ for a retention of around $2/3$ of the signal level \cite{lambrick_ray_2018}. Second, we require sufficient space between the pinhole and the scattering plane, $F$, for bulk samples and straightforward sample manipulation. For a target beamwidth of $\Phi\sim\SI{300}{\nano\metre}$ this leads, after correcting for the $\theta=\ang{45}$ geometry, to a lower limit of roughly $f>\SI{650}{\micro\metre}$. Imposing $F\sim\SI{500}{\micro\metre}$ then gives a practical design value of $f\sim\SI{710}{\micro\metre}$.
Applying constrained optimisation to equation \ref{eq:beamwidth_use} in the same manner as Bergin (details in \ref{sec:non-normal_incidence}) gives the optimum pinhole diameter and angular source size as
\begin{align}
	d_0 &= \sqrt{6} \, \phi_x\label{eq:optimised1}\\
	\beta_0 &= \frac{\sqrt{3}}{\sqrt{2}f}\left(\frac{\phi_x^2}{2} - \frac{a^2}{6\phi_x^2}\right)^\half.\label{eq:optimised2}
\end{align}
Here $a=0.42\lambda d/\cos^2\theta$, with $\lambda = \SI{0.57}{\angstrom}$ for room-temperature helium.

For a target intrinsic beam \textsc{fwhm} of approximately \SI{300}{\nano\metre}, the constrained optimisation yields the design targets summarised in table~\ref{tab:system}, including an optimum pinhole diameter of order \SI{440}{\nano\metre} and an optimum angular source size of order $2.2\times10^{-4}$ rad. The fact that the optimised pinhole diameter exceeds the target intrinsic beamwidth is not contradictory: the optimisation is carried out in terms of the projected beamwidth along one dimension and the intensity from a circular aperture projected onto one dimension will not form a top hat distribution but a quadratic intensity profile with a FWHM smaller than the pinhole diameter.

For the chosen working distance, the optimised design predicts sub-micron performance, with geometric, source-size and diffraction terms all contributing materially to the overall beamwidth rather than the resolution being dominated almost entirely by geometric broadening (table~\ref{tab:system}).

The optimisation also implies a substantial signal penalty relative to the earlier large-working-distance configuration. Reducing the pinhole diameter and increasing the source--pinhole distance both reduce the transmitted flux, while the final signal level also depends on the detector aperture and contrast mechanism. For this reason the signal cannot be predicted from geometry alone, and in practice it is more meaningful to estimate it by comparison with a known experimental configuration using the same source and detector\cite{lambrick_multiple_2020,sam_m_lambrick_formation_2021}. Relative to that earlier configuration, the optimised design is expected to incur a large raw signal loss, of approximately $\times 18$ before mitigation. However, a significant fraction of that loss can be mitigated by detector-aperture redesign discussed in section \ref{sec:pinhole-plate}.

\begin{table}[h]
    \centering
    \begin{tabular}{r|l}
        Quantity & Value \\
        \hline
        & \\
        Target intrinsic beamwidth $\Phi$ & \SI{300}{\nano\metre} \\
        Target projected beamwidth $\Phi_x$ & \SI{424}{\nano\metre} \\
        Optimised pinhole diameter $d_0$ & \SI{440}{\nano\metre} \\
        Optimised angular source size $\beta_0$ & $2.20\times10^{-4}\,$rad \\
        Design working distance $f$ & \SI{0.71}{\milli\metre} \\
        Design pinhole--sample spacing $F$ & \SI{0.50}{\milli\metre} \\
        Helium wavelength $\lambda$ & \SI{0.57}{\angstrom} \\
        Manufactured pinhole diameter $d_0$ & \SI{470}{\nano\metre} \\
        Predicted intrinsic beamwidth $\Phi$ & \SI{330}{\nano\metre} \\
        Predicted projected beamwidth $\Phi_x$ & \SI{470}{\nano\metre} \\
        Geometric contribution & 68\% \\
        Source-size contribution & 64\% \\
        Diffraction contribution & 36\% \\
        Estimated raw signal penalty & $\sim\times18$ \\
    \end{tabular}
    \caption{Optimised design targets and predicted performance for the high-resolution large-working-distance configuration. The intrinsic beamwidth is $\sqrt{2}$ smaller than the projected beamwidth owing to the $\ang{45}$ incidence geometry. Contribution percentages are expressed relative to the overall predicted beamwidth. The predicted beamwidth and contributions are based on the actual diameter of the manufactured pinhole.}
    \label{tab:system}
\end{table}

\section{Experimental Implementation and Design}

The optimisation of Section~\ref{sec:optical_optimisation} specifies the atom-optical requirements for sub-micron, large-working-distance operation, but realising that design in practice raises three distinct implementation challenges:
\begin{enumerate}
    \item increasing the source--pinhole distance to reduce the angular source size,
    \item reducing the working distance without losing practical sample access, and
    \item compacting the local scattering geometry sufficiently to preserve useful detector acceptance despite the smaller pinhole--sample separation.
\end{enumerate}

The experimental redesign was therefore organised around three corresponding hardware changes: a chamber geometry that allows the source distance to be varied over a larger range, a reduced working-distance sample region, and a miniaturised pinhole-plate that implements the compact pinhole--sample--detector geometry required by the optimisation while remaining firmly in the advantageous Cambridge-type geometry regime and defining a meaningful coincidence point. To enact these changes, a new sample chamber was designed and installed, a cross-sectional render of which is shown in figure~\ref{fig:A-SHeM}, together with a newly designed defining optical element, the high-resolution pinhole-plate. 

\subsection{New sample chamber}

\begin{figure}
	\centering
	\includegraphics[width=\linewidth]{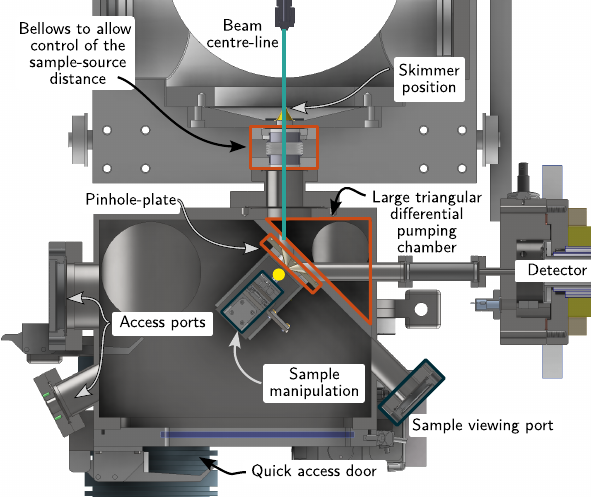}
	\caption{A cross-sectional render of the Cambridge SHeM, with a new sample chamber installed: the sample chamber is the large rectangular chamber in the lower left of the render. Key features of the new chamber are highlighted. The quick access door and the large number of access ports (3 more on the top of the chamber are not rendered) are usability improvements. The large triangular differential pumping chamber decreases background, while the adjustable bellows, pinhole-plate, and viewing port enable the optical system changes to enable the high resolution imaging mode. The sample is mounted at the yellow dot in the centre of the render.}
	\label{fig:A-SHeM}
\end{figure}

The new rectangular chamber is split into two sub-chambers which maximises signal to noise through the reduction of background by an optimisation of the pumping speed in the differential chamber\cite{fahy_highly_2015}. The chamber is designed to have ample space for extended sample manipulation, alternative pinhole-plate designs (such as that used by Radi\'{c} et al.\cite{radic_3d_2024}), and other new sample environments, with a generous number of access ports on the side and top giving direct access to the sample area. 
The angular source size can be manipulated by changing the distance between the skimmer, which defines the source size, and the pinhole using the bellows highlighted in figure \ref{fig:A-SHeM}. A large range of source distances are possible, from a minimum of $\sim\SI{130}{\milli\metre}$, to $\sim\SI{400}{\milli\metre}$. The same $\SI{10}{\micro\metre}$ nozzle and $\SI{100}{\micro\metre}$ skimmer we used as the original Cambridge SHeM, meaning the virtual source size, $\SI{80}{\micro\metre}$ FWHM at nozzle pressures of $\approx\SI{75}{bar}$, is the same as previously reported\cite{bergin_instrumentation_2018}.
For the high-resolution configuration reported here, the source--pinhole distance was set to $L=385\pm\SI{20}{\milli\metre}$, corresponding to a realised angular source size of $\beta=2.18\pm0.11\times10^{-4}\,$rad. This places the implemented source geometry in close agreement with the optimised requirement identified in Section~\ref{sec:optical_optimisation}.

\subsection{New pinhole-plate optical element}
\label{sec:pinhole-plate}

\begin{figure}
	\centering
	\includegraphics[width=\linewidth]{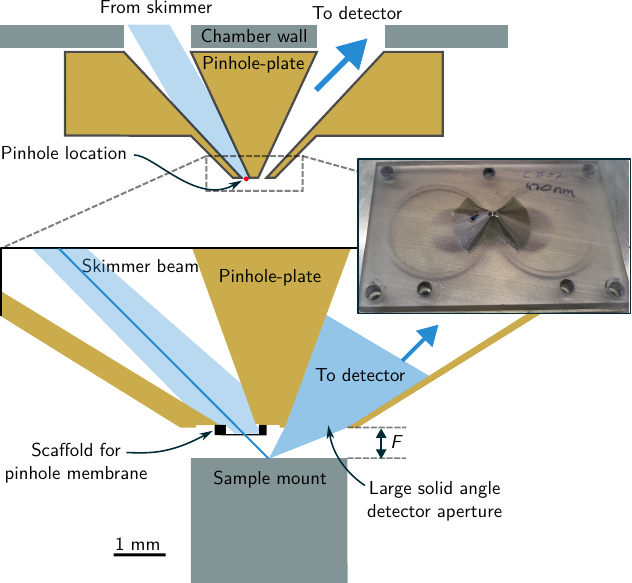}
	\caption{Top, schematic of a pinhole plate (orange) in the Cambridge SHeM, the pinhole plate is mounted to the sample chamber wall and separates the incoming source chambers from the detector chamber, it defines the path of the atoms from the source to the pinhole and from the detector aperture to the detector chamber. Bottom, a to-scale cross-sectional view of the tip of the high-resolution pinhole-plate, showing the scaffold for the pinhole membrane and the large solid angle detector aperture. The inset shows the realised 3D-printed assembly with the pinhole membrane installed.}
	\label{fig:pinhole_plate}
\end{figure}

The pinhole-plate is the defining optical element of the Cambridge-type SHeM in the immediate sample region\cite{bergin_complex_2021}. The top panel of figure \ref{fig:pinhole_plate} shows a cross section of a pinhole plate, it consists of a mount for a silicon nitride membrane into which is milled the pinhole, a detector aperture which collects the scattered atoms, as well as pathways from the differential source and to the detector. The pinhole membrane is mounted onto the pinhole plate parallel to the sample, and hence $\ang{45}$ to the incident beam, which minimises the risk of interference between the sample and the membrane mounting scaffold (overall dimensions of the scaffold are similar to the perpendicular working distance $F$ in both the previous and current designs). The pinhole plate mounts to the wall of the sample chamber and isolates the source side of the instrument from the detector side of the instrument. In terms of the atom optics it fixes the local scattering geometry of the instrument: it sets the working distance, determines the solid angle subtended by the detector aperture, and establishes the relative positions of the pinhole, sample and detector. To realise our optimised system the pinhole and detector aperture need to be brought closer together, while simultaneously increasing the solid angle of the detector aperture. Therefore  a miniaturisation of the pinhole plate features was required. A cross-sectional diagram of a generic pinhole plate and a to scale cross-section of the defining apertures of the new high-resolution pinhole-plate, manufactured by 3D printing in plastic\cite{radic_application_2025}, is shown in figure~\ref{fig:pinhole_plate}.  
With the reduced working distance of $f\approx\SI{0.71}{\milli\metre}$ the pinhole position becomes close to the centre of the detector aperture, $\sim\SI{1.5}{\milli\metre}$, with the detector aperture also enlarged relative to the previous design. The new working distance is approximately one quarter of that used previously with the Cambridge SHeM, although still around one to one-and-a-half orders of magnitude larger than the short-working-distance Portland-type\cite{witham_increased_2012}.

\begin{figure}
	\centering
	\includegraphics[width=0.5\linewidth]{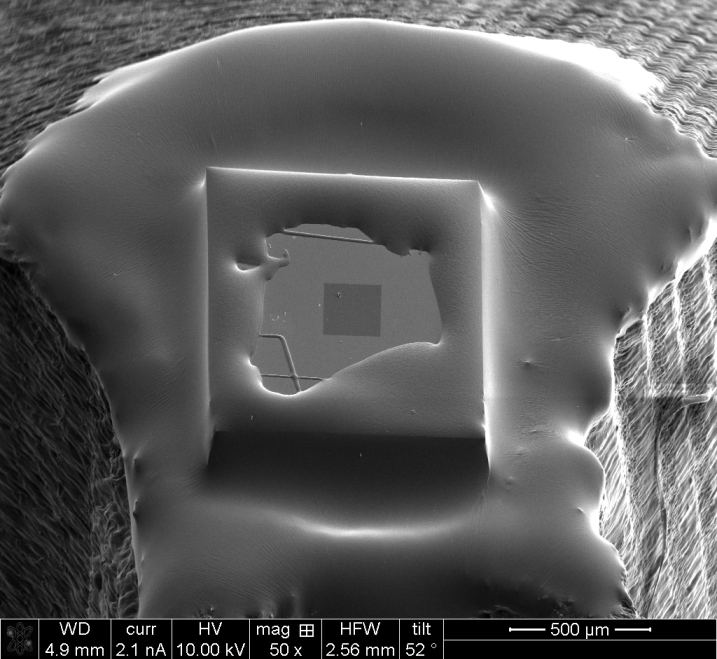}%
	\includegraphics[width=0.5\linewidth]{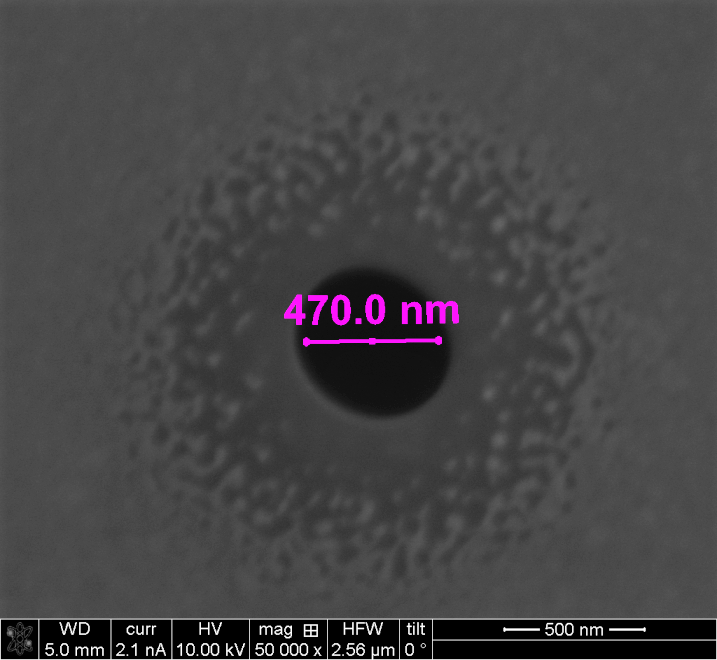}%
	\caption{SEM micrographs of the FIB-milled pinhole in the Si--N membrane. The custom \SI{1}{\milli\metre} square membrane scaffold is visible, demonstrating the miniaturised support structure required by the compact pinhole-plate geometry. The membrane is mounted into the pinhole-plate prior to FIB milling.}
	\label{fig:pinhole}
\end{figure}

The miniaturised design made the membrane scaffold itself a critical design constraint. Commercially available pinholes, or the standard Agar Scientific membranes used previously\footnote{Agar Scientific Ltd. silicon nitride membranes in support frames compatible with 3mm TEM holders: \url{https://www.agarscientific.com/silicon-nitride-membranes-200-181-m-substrate-thickness}.}, have lateral dimensions of order \SI{3}{\milli\metre}. These would occupy too much of the compact sample region to leave sufficient space for the detector aperture. For this reason a custom square silicon nitride membrane, with scaffold approximately \SI{1}{\milli\metre} wide was used, allowing a corresponding miniaturisation of the pinhole-plate assembly. The new membrane scaffold is shown to scale in figure \ref{fig:pinhole_plate} and final milled structure are shown directly in the SEM micrographs of figure~\ref{fig:pinhole}. After mounting the membrane onto the pinhole-plate with vacuum leak sealant, the assembly was sputter-coated with gold and transferred to a FIB-SEM\footnote{FEI Helios NanoLab 600i with a Ga\textsuperscript{+} beam.} for focused ion beam milling of the final pinhole. The realised pinhole diameter after milling was $d=\SI{470}{\nano\metre}$, slightly larger than the optimised target value of approximately \SI{440}{\nano\metre} but still within the regime required for sub-micron performance. As seen from the right hand micrograph in figure \ref{fig:pinhole} the pinhole is well approximated by a circle, meaning our assumption of a circular aperture in equation \ref{eq:beamwidth_use} remains reasonable. Finite roughness (induced by the FIB milling) below the resolution of the SEM micrograph near the edge of the aperture may be present and could result in some intensity being scattered more broadly. However, such imperfections will only scatter a small fraction of the flux as the overall shape of the aperture remains as intended, and, given the lengthscale, will largely manifest as a slightly elevated background signal with no direct impact on the primary beam spot size. The thickness of the membrane used was $\SI{50}{\nano\metre}$, ensuring scattering from the side walls of the aperture (which will occur due to the $\ang{45}$ incidence angle) remains an order of magnitude less than the flux passed straight through. 

The enlarged detector aperture in the pinhole-plate redesign was also used to recover as much contrast-bearing signal as possible after the substantial signal losses imposed by the smaller pinhole and increased source--pinhole distance. In the new design, the solid angle subtended by the detector aperture, $\Omega$, is increased from $\SI{0.094}{sr}$ in the previous configuration to $\SI{1.44}{sr}$, an increase by a factor of approximately 15. For the diffuse topographic contrast mode used here, previous work has shown that increasing detector solid angle can improve contrast-to-noise performance without a corresponding penalty in image quality, even though a larger fraction of the scattering distribution is averaged\cite{sam_m_lambrick_formation_2021}. For an increase in solid angle of $\times 15$, and a detector aperture occupying a circular region of solid angle, the increase in the ratio of contrast to noise is found to be equivalent to that achieved by a raw increase of single of a factor $\times 8$ (see chapter 3 Lambrick 2021\cite{sam_m_lambrick_formation_2021}).  Therefore, in utilising a larger detector aperture we have effectively compensated for some of the predicted raw signal penalty by $\times8$.

The enlarged detector acceptance, however, does not fully compensate the raw $\times 18$ signal losses associated with the reduction in pinhole diameter and the increased source--pinhole distance, but it does remain practical. Thus, the net effective  reduction in signal is approximately $18/8=2.25$ relative to the earlier configuration, but this was found to be acceptable in practice through either a moderately increased measurement time or acceptance of a slightly degraded SNR. The use of a larger aperture to compensate for a smaller beam is, however, contrast-mechanism specific: very large apertures still produce good diffuse topographic contrast\cite{lambrick_observation_2022}, which is the imaging mode most relevant to the samples presented here. In particular, the resulting design is less well suited to forms of contrast that depend more strongly on angular selectivity, such as diffraction contrast\cite{bergin_observation_2020,von_jeinsen_2d_2023}, the measurement of which is a major advantage of the Cambridge-type instrument. The pinhole-plate used in the current work has a very large angular acceptance, $\sim\ang{40}$ in the scattering plane, detecting atoms with $\Delta K$ of a range $\sim 4\AA^{-1}$ when centred on specular, and thus would almost always capture multiple diffraction peaks. Therefore, spot profile diffraction could not be measured with the high-resolution arrangement present here ``as is''. However, the optimisation of the incident optics could be combined with a new pinhole plate and appropriate consideration of the detected diffraction signal level, to realise sub-micron resolution for diffraction measurements.

\subsection{Alignment}

An advantage of the Cambridge-type geometry, and of pinhole SHeM more generally, is the surprisingly relaxed alignment tolerances, even for a high resolution instrument such as that presented in the current work. Primary pinhole is between three apertures, as seen in figure \ref{fig:optical_system}, the nozzle, the skimmer, and the pinhole. The skimmer is $\sim\SI{100}{\micro\metre}$ and in a fixed position while the effective size of the nozzle (virtual source size) is also $\sim\SI{100}{\micro\metre}$. Alignment between the skimmer and nozzle is done though a micrometer stage holding the nozzle. The pinhole is on the nanoscale, but by the time the incident beam reaches the pinhole position it is a few millimetres wide, thus positioning tolerance of the pinhole is again a matter of large fractions of a millimetre, and not fractions of a micron. In addition, post milling of the pinhole, as discussed in section \ref{sec:pinhole-plate}, after mounting the membrane in the pinhole-plate allows pricse positioning of the pinhole within the overall pinhole-plate. For imaging the sample merely neads to be beneath the incident beam. Thus, initial sample alignment, and measurement of the working distance, can be performed with an optical camera ex vacuo through a vacuum window\footnote{Raspberry pi HQ 12MP camera was used in the current work. \url{https://www.raspberrypi.com/products/raspberry-pi-high-quality-camera/}}, wile fine alignment is achieved by acquiring a low resolution overview SHeM micrograph and the aid of the encoded piezo-electric sample positions which provide an absolute coordinate system.

\section{Measured beamwidth}

\begin{figure}
    \centering
    \includegraphics[width=0.75\linewidth]{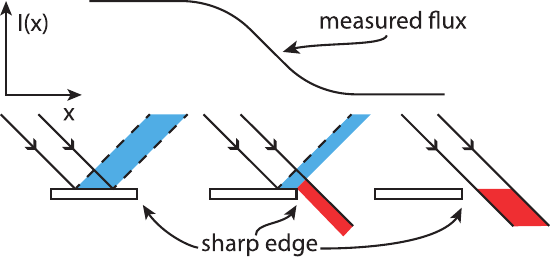}
    \caption{Scanning across a step edge with steps smaller than the beam width will result in the detected helium flux, $I(x)$ shown in blue, changing as a greater fraction of the beam falls off the edge, in red.}
    \label{fig:step_edge}
    \vspace{\baselineskip}
    \includegraphics[width=\linewidth]{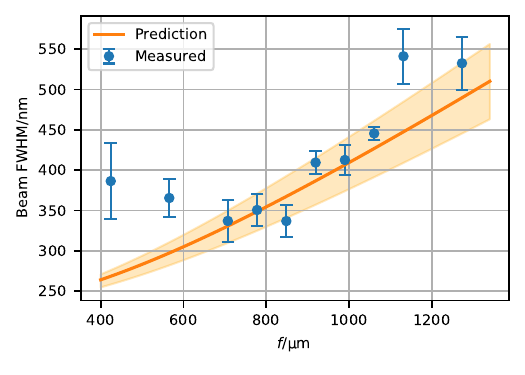}
    \caption{Beamwidth measurements, plotted as the intrinsic beamwidth $\phi$, taken across a wire edge provided by a copper TEM grid as a function of the working distance $f$. Two different edges were used for each measurement and a weighted average taken. The predicted beamwidth uses equation \ref{eq:beamwidth_use} and the confidence interval represents one standard deviation in the predicted value based on the uncertainty in the source size $s$\cite{bergin_instrumentation_2018}. Working distances are all measured to $\pm\SI{25}{\micro\metre}$. }
    \label{fig:resolution_measurements}
\end{figure}

To measure the beamwidth, we scan over a sharp edge on a standard copper TEM grid\footnote{Square Pattern 50 Mesh TEM Support Grids, from Agar Scientific \url{www.agarscientific.com/tem/grids-agar/square-50-mesh-tem-support-grids}.} and monitor the detected helium signal, as demonstrated in figure \ref{fig:step_edge}. Multiple line scans (4-6) were acquired in a rectangular box, see figure \ref{fig:line_scan_image}, and the resulting beamwidths averaged to improve precision, step sizes of \SI{10}{\nano\metre} were used along the scan direction. The beamwidth was extracted from the measured signal by fitting an error function and extracting the associated standard deviation. Figure \ref{fig:resolution_measurements} shows the measured intrinsic beamwidths plotted as a function of the working distance, as well as a predicted beamwidth from equation \ref{eq:beamwidth_use}. The measured beamwidths follow the expected trend, with beamwidths near and slightly larger than the design working distance agreeing with the prediction. The smallest measured beamwidth was $340\pm\SI{20}{\nano\metre}$ at a working distance of $f=850\pm\SI{25}{\micro\metre}$. The corresponding beamwidth in the scanning plane is $\Phi_x=480\pm\SI{30}{\nano\metre}$, and the predicted beamwidth in the other scanning direction is $\Phi_y=\SI{390}{\nano\metre}$. The effective resolution achieved will depend on the orientation of the topographic feature of interest: a feature perpendicular to the beam will encounter an effective beamwidth of $\SI{340}{\nano\metre}$, while one in the scanning plane will encounter an effective beamwidth of $\SI{480}{\nano\metre}$ along $x$, with many real world feature lying in between. In the interpretation of micrographs such variation is intuitively accounted for by considering the effective `angle of view' of the instrument being $\ang{45}$ to the scanning plane parallel to the incident beam, similar to other scanning beam techniques.

We can see from figure \ref{fig:resolution_measurements} that we maintain a narrow beamwidth over a significant range of working distances, $\sim550-\SI{1100}{\micro\metre}$, satisfying our requirement for a large depth of field. For working distances significantly smaller than the design distance, we see divergence from the expected beamwidth, for these data points, the signal to noise was low yielding poor fits; therefore we attribute the divergence to decreased signal to noise when the instrument is used away from its design geometry. We confirm this interpretation by estimating the signal to noise ratio (SNR) as a function of the working distance from the same set of scans over step edges, results shown in \ref{app:SNR}, and show a broad peak in SNR between $\sim\SI{500}{\micro\metre}$ through $\sim\SI{1100}{\micro\metre}$. These observations are consistent with our previous results that demonstrated an `effective depth of field' for pinhole SHeM of $\sim\pm f/3$\cite{lambrick_ray_2018}. 

\section{High resolution imaging}

\begin{figure}
    \centering
    \includegraphics[width=\linewidth]{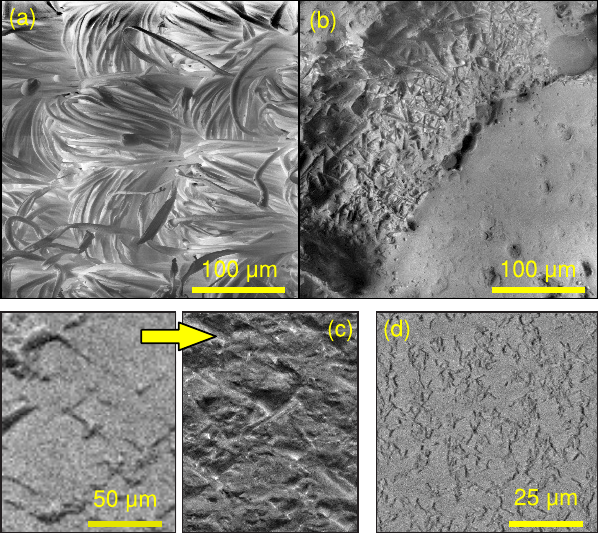}
    \caption{Example high-resolution SHeM micrographs demonstrating a large depth of field (top row) and high fidelity (bottom row). (a) Reusable PPE face mask fabric provided by L. Fruk (Chemical Engineering Department, University of Cambridge). (b) Formaldehyde-fixed \textit{Pseudomonas aeruginosa} bacterial biofilm on a generic glass slide. (c) Polycrystalline diamond from Element Six, eroded by T. Cowie with \SI{100}{\micro\metre} silicon carbide particles \cite{henderson_solid_2020}. The left panel shows the surface prior to resolution improvements; the right panel (c) shows the significant increase in surface detail following the improvements. (d) Isolated \textit{Clostridioides difficile} (C. diff) bacteria provided by L. Dawson (London School of Hygiene and Tropical Medicine, LSHTM) \cite{dawson_extracellular_2021}, grown directly on round polymer coverslips (Thermo Fisher 174950). Panels (a) and (b) were acquired with a pixel size of \SI{1.00}{\micro\metre} and count time per pixel of \SI{750}{\milli\second}, while panels (c) (right) and (d) were acquired with a pixel size of \SI{0.50}{\micro\metre} and count time of \SI{1500}{\milli\second} and \SI{1800}{\milli\second} respectively. The image in the old configuration in panel (c) was acquired with a pixel size of \SI{1.5}{\micro\metre} and a count time of \SI{1500}{\micro\metre}.}
    \label{fig:hi-res-images}
\end{figure}

Four example samples were chosen to explore the capabilities of the new high resolution imaging mode, these are all presented in figure \ref{fig:hi-res-images}. SHeM micrographs in the high resolution configuration were acquired with a count time per pixel of between \SI{750}{\milli\second} and \SI{1800}{\milli\second}, which is within the range from the count times used in the previous configuration, for example Lambrick et al. 2020\cite{lambrick_multiple_2020} used \SI{750}{\milli\second} and the eroded diamond image on the right panel of figure \ref{fig:hi-res-images} (c) used \SI{1500}{\milli\second}. Thus we demonstrating that our improvement in beamwidth is practical and remains usable. All micrographs in figure \ref{fig:hi-res-images} exhibit sufficient signal-to-noise ratio for clear topographic imaging, with features readily distinguishable above the noise floor.

Figure  \ref{fig:hi-res-images} (b) shows a micrograph of \textit{Pseudomonas aeruginosa} which has formed a bacterial biofilm and exhibits very high topographic detail, with strong contrast. In particular, the complex pattern of the formaldehyde crystals, recognised by their string-like appearance, easily made out. 
Figure \ref{fig:hi-res-images} (d) shows a micrograph of isolated \textit{Clostridioides difficile} (\textit{C. diff}) bacteria, these appear as small rod shapes, and are $\lesssim\SI{1}{\micro\metre}$ in diameter, in the micrograph it is the dark masks cast by the bacteria that dominate contrast, with the smallest $\sim\SI{500}{\nano\metre}$ across. It would not have been possible to resolve the individual bacteria with our previous configuration -- or any other large working distance pinhole helium microscope\cite{vonJeinsen2026SurfaceMicroscopy}. Both bacterial specimens were fixed with formaldehyde. 

Figure \ref{fig:hi-res-images} (c) shows micrographs of mechanically eroded synthetic diamond imaged with the old configuration (left) and the new optimised configuration (right). The eroded diamond has a strongly faceted surface arising from chunks of the material being removed during the erosion process. The micrograph with the previous setup could only resolve some of the largest grooves, $>\SI{2}{\micro\metre}$ across, while the optimised configuration shows extensive detail of smaller grooves, some $<\SI{1}{\micro\metre}$ across, between the larger ones. 

The final sample was a reusable PPE face mask fabric and presents a particularly difficult structure to image due to being both insulating and having features as large perpendicular to the image plane as within it. The SHeM micrograph in figure \ref{fig:hi-res-images} (a) resolves the fabric filaments deep into the material as well as at the surface, demonstrating that we have been successful in maintaining the impressive depth of field of the Cambridge-type SHeM in the sub-micron beamwidth regime. 

\section{Outlook}

The present work describes the optimisation of and experimental advances in large-working-distance pinhole SHeM as a viable sub-micron microscopy platform. The achieved intrinsic beamwidth of \SI{340}{\nano\metre} narrows the gap between the Cambridge-type geometry and the best reported short-working-distance Portland-type performance, while retaining the advantages of larger working distance, including a large depth of field, tolerance of significant sample topography, and the modular framework of the Cambridge SHeM. Thus, large-working-distance SHeM can now operate in a resolution regime that begins to make fuller use of its non-destructive and surface-sensitive character.

An important consequence of the present optimisation is that the beamwidth is no longer dominated almost entirely by simple geometric broadening. Instead, geometric, and source-size contributions are now very similar, indicating an optimised regime and that further gains from passive optimisation of the same architecture are likely to be limited. The diffraction term remains the smallest, but cannot uniquely be tuned due to the fixed helium wavelength. Some additional improvement may still be possible by trading working distance against resolution, for example by moving towards normal incidence or reducing the pinhole--sample separation further, but such changes would come at the cost of depth of field and reduced tolerance of sample topography. In practice such changes would involve abandoning the advantages of the modular large-working distance Cambridge-type in favour of a more specialised hybrid type with short working distances and a skimmed beam. Therefore, the present work represents not only an improvement in beamwidth, but it clarifies the current technical frontier. Within the pinhole-collimated geometry, further substantial gains are likely to require a complete rethinking of the instrumental design, rather than further evolution of the current design paradigm. In particular, active focusing approaches such as Fresnel zone plates remain the most promising route if helium microscopy is to move substantially beyond the performance demonstrated here\cite{flatabo_reflection_2023}. The present results therefore help define the performance envelope of optimised large-working-distance pinhole SHeM and provide a useful benchmark against which future focused-beam approaches can be assessed.

A promising advance on the current work will be to combine sub-micron resolution with the atom-diffraction capabilities recently demonstrated in the literature\cite{von_jeinsen_2d_2023,hatchwell_measuring_2024}. The optimisation of the incident optics used here can directly be translated to such measurement, however, some further optimisation will be needed on the detection optics as that is currently optimised for diffuse topographic contrast.

The sub-micron regime already opens up a much broader range of measurements than was previously accessible to large-working-distance SHeM. For topographic imaging, it extends the usefulness of the technique for delicate and insulating specimens whose relevant structure lies below the micron scale. For atom-diffraction and spot-profile measurements, it makes it more realistic to probe smaller individual domains and heterogeneous regions, including exfoliated and spatially non-uniform 2D systems, for which micron-scale beam sizes were previously a substantial limitation\cite{von_jeinsen_2d_2023,radic_defect_2024}. More broadly, the work shows that the practical advantages of the Cambridge-type geometry need not be confined to low-resolution operation, and that helium microscopy can increasingly be considered not simply as a specialised surface-science instrument, but as a flexible microscopy platform with growing relevance for challenging real-world specimens.
\section*{Acknowledgements}

\noindent The work was supported by \textsc{epsrc} grants EP/R008272/1 and EP/R008051/1. The work was performed in part at \textsc{corde}, the Collaborative R\&D Environment established to provide access to physics related facilities at the Cavendish Laboratory, University of Cambridge and \textsc{epsrc} award EP/T00634X/1. SML acknowledges support from IAA award EP/X525686/1. The authors acknowledge support from Ionoptika Ltd. We thank Dr L Dawson of the London School for Tropical Medicine for the biofilm samples, T Cowie and Element Six for the eroded diamond samples, and Dr L. Fruk from the University of Cambridge Chemical Engineering department for the fabric face mask sample.

\section*{Data availability}

Raw and processed data to accompany this work are available at \url{doi.org/10.5281/zenodo.22125868}.

\bibliography{references.bib}

\clearpage

\appendix

\section{Constrained optimisation for non-normal incidence}
\label{sec:non-normal_incidence}

\begin{figure}
    \centering
    \includegraphics[width=0.9\linewidth]{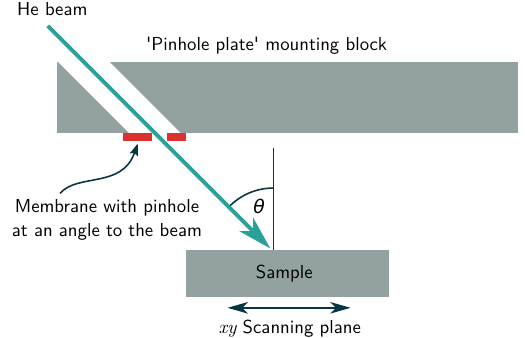}
    \caption{In the Cambridge SHeM the pinhole is mounted on a `pinhole plate'. The pinhole plate is parallel to the scanning plane of the sample manipulators, and therefore for non-zero $\theta$ the pinhole is at an angle to the incident helium beam.}
    \label{fig:pinhole_position}
\end{figure}

The formula for the beam standard deviation, $\phi$, derived by Bergin is
\begin{equation}\label{eq:bergin}
    \phi^2 = \underbrace{\left(\frac{d}{2\sqrt{3}}\right)^2}_{\text{geometric}} + \underbrace{\left(\frac{\beta f}{\sqrt{3}}\right)^2}_{\text{source size}} + \underbrace{\left(\frac{0.42\lambda f}{d}\right)^2}_{\text{diffraction}},
\end{equation}
where $d$ is the pinhole diameter, $f$ is the working distance, $\beta$ is the angular source size and $\lambda$ is the wavelength of the helium atoms. Modifications are required in order to apply equation \ref{eq:bergin} to a non-normal incidence configuration, such as the Cambridge SHeM. We must also consider that the pinhole in the Cambridge SHeM is mounted parallel to the $xy$ scanning axes, and therefore at an angle to the incident beam, as shown in figure \ref{fig:pinhole_position}. With non-normal incidence the beamwidth may be different in the two principle scanning axes, we consider the horizontal scanning axis, that which is in plane with the incoming helium beam and follows the arrow in figure \ref{fig:pinhole_position}. We will consider the three terms in equation \ref{eq:bergin} with respect to measuring a beamwidth projected into the horizontal scanning axis, which we shall label as the $x$-axis.
\begin{labeling}{\sffamily geometric}
\item[\textsf{geometric}] As the pinhole is in the same plane as the scanning axis of interest, then the geometric projection of the pinhole is unchanged.
    \item[\textsf{source}] A beamwidth measured parallel to the scanning axis will be observed to be $1/\cos\theta$ larger than it actually is.
    \item[\textsf{diffraction}] As with the source size, the angle between the beam and the scanning axis will introduce a factor of $1/\cos\theta$. In addition, because the beam is arriving at an angle to the pinhole, the effective width of the pinhole is reduced by a factor $\cos\theta$. 
\end{labeling}
Thus the beam standard deviation as measured in the scanning plane, $\phi_x$, is
\begin{equation}
    \phi_x^2 = \left(\frac{d}{2\sqrt{3}}\right)^2 + \left(\frac{1}{\cos\theta}\frac{\beta f}{\sqrt{3}}\right)^2 + \left(\frac{1}{\cos\theta} \frac{0.42\lambda f}{d\cos\theta}\right)^2.
\end{equation}

Define the constant $a=0.42\lambda f/\cos^2\theta$. Following the formulation by Bergin the Lagrange multiplier is
\begin{equation}
    \mathcal{L} = \gamma d^2\beta^2 - \Lambda \left[ \sqrt{\left(\frac{d}{2\sqrt{3}}\right)^2 + \left(\frac{\beta f}{\sqrt{3}\cos\theta}\right)^2 + \left(\frac{a}{d}\right)^2} - \phi_x\right],
\end{equation}
which must be minimised.
\begin{align}
	\frac{\partial \mathcal{L}}{\partial \beta} &= 2\gamma d^2 \beta - \frac{\Lambda}{\phi_x}\frac{2\beta f^2}{3} \\
	\implies\;\frac{\Lambda}{\phi_x} &= \frac{3\gamma d_0^2}{f^2} \\
	\frac{\partial \mathcal{L}}{\partial d} &= 2\gamma \beta^2 d - \frac{\Lambda}{\phi_x}\left( \frac{d}{12} - \frac{a^2}{d^3} \right)
\end{align}
as for normal incidence.
\begin{align}
	\frac{3\gamma d_0^2}{f^2} &= \frac{2\gamma \beta_0^2 d_0^4}{\frac{d_0^4}{12} - a^2} \\
	\implies \;\; \frac{2f^2\beta_0^2}{3} &= \frac{d_0^2}{12} - \frac{a^2}{d_0^2}
\end{align}
Which gives the constrained optimisation values for pinhole size and virtual source size as
\begin{align}
	d_0 &= \sqrt{6} \phi_x\\
	\beta_0 &= \frac{\sqrt{3}}{\sqrt{2}f}\left(\frac{\phi_x^2}{2} - \frac{a^2}{6\phi_x^2}\right)^\half
\end{align}
which is the same as for normal incidence with a re-definition of the constant \(a\). 

\section{Extra figures}

\begin{figure}[h]
	\centering
	\includegraphics[width=0.65\linewidth]{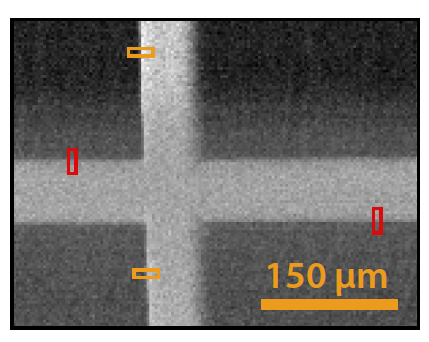}
	\caption{SHeM micrograph of a standard TEM grid, with the two location of the line scans performed to measure resolution highlighted in yellow. Two different locations were used to ensure reliable results, and at each locations a small number (4-6) scans were acquired and averaged. The data for each was fitted independently to an error function and the results averaged. The micrograph was acquired with a count time of \SI{705}{\milli\second} per pixel and a pixel size of \SI{3}{\micro\metre}.}
	\label{fig:line_scan_image}
\end{figure}

\section{SNR with $f$}
\label{app:SNR}

\begin{figure}[h]
    \centering
    \includegraphics[width=\linewidth]{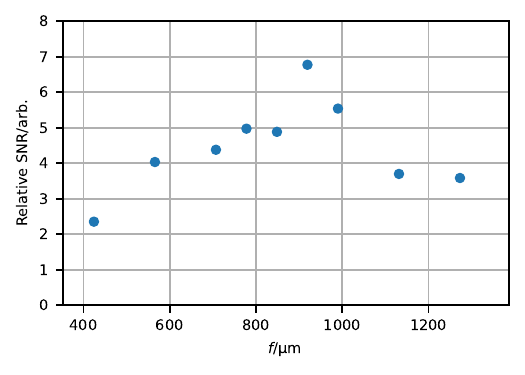}
    \caption{Estimations of how the SNR varies with the working distance $f$. The datapoint for $f=\SI{1060}{\micro\metre}$ was found to have a significantly higher SNR than all other data, this is attributed to detector behaviour and that point has been excluded from the plot.}
    \label{fig:SNR}
\end{figure}

The effective depth of field in SHeM is determined by a combination of the beamwidth and the achievable signal to noise ratio for a specified working distance. We estimate the SNR by taking the ratio between the residuals of the fit and the magnitude of the intensity drop across the step edge. These are then normalised by the scan time. Given differing imaging conditions and potential different detector behaviour, the estimated values of SNR are relative only, with the absolute value not meaningful. In addition, identical detector behaviour cannot be guaranteed for measurements taken multiple days apart so comparisons remain qualitative. The estimate relative SNR is plotted in figure \ref{fig:SNR}, and shows a broad maximum between $\sim\SI{600}{\micro\metre}$ and $\sim\SI{1000}{\micro\metre}$, representing the ``effective depth of field'' due to SNR.

\end{document}

%% file: optical_system.tex
\begingroup%
  \makeatletter%
  \providecommand\color[2][]{%
    \errmessage{(Inkscape) Color is used for the text in Inkscape, but the package 'color.sty' is not loaded}%
    \renewcommand\color[2][]{}%
  }%
  \providecommand\transparent[1]{%
    \errmessage{(Inkscape) Transparency is used (non-zero) for the text in Inkscape, but the package 'transparent.sty' is not loaded}%
    \renewcommand\transparent[1]{}%
  }%
  \providecommand\rotatebox[2]{#2}%
  \newcommand*\fsize{\dimexpr\f@size pt\relax}%
  \newcommand*\lineheight[1]{\fontsize{\fsize}{#1\fsize}\selectfont}%
  \ifx\svgwidth\undefined%
    \setlength{\unitlength}{252bp}%
    \ifx\svgscale\undefined%
      \relax%
    \else%
      \setlength{\unitlength}{\unitlength * \real{\svgscale}}%
    \fi%
  \else%
    \setlength{\unitlength}{\svgwidth}%
  \fi%
  \global\let\svgwidth\undefined%
  \global\let\svgscale\undefined%
  \makeatother%
  \begin{picture}(1,0.5033302)%
    \lineheight{1}%
    \setlength\tabcolsep{0pt}%
    \put(0,0){\includegraphics[width=\unitlength,page=1]{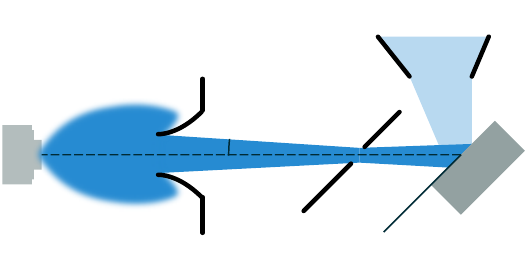}}%
    \put(0.46480348,0.24694195){\color[rgb]{0,0,0}\makebox(0,0)[t]{\lineheight{1.25}\smash{\begin{tabular}[t]{c}$\beta$\end{tabular}}}}%
    \put(0.91634091,0.18605956){\color[rgb]{0,0,0}\makebox(0,0)[t]{\lineheight{1.25}\smash{\begin{tabular}[t]{c}$\phi_x$\end{tabular}}}}%
    \put(0.78683293,0.1566966){\color[rgb]{0,0,0}\makebox(0,0)[t]{\lineheight{1.25}\smash{\begin{tabular}[t]{c}$\pi-\theta$\end{tabular}}}}%
    \put(0.2575857,0.21976758){\color[rgb]{0,0,0}\makebox(0,0)[t]{\lineheight{1.25}\smash{\begin{tabular}[t]{c}$\sigma$\end{tabular}}}}%
    \put(0.38690487,0.37402364){\color[rgb]{0,0,0}\makebox(0,0)[t]{\lineheight{1.25}\smash{\begin{tabular}[t]{c}Skimmer\end{tabular}}}}%
    \put(0.05831149,0.28414199){\color[rgb]{0,0,0}\makebox(0,0)[t]{\lineheight{1.25}\smash{\begin{tabular}[t]{c}Nozzle\end{tabular}}}}%
    \put(0.60588085,0.38518693){\color[rgb]{0,0,0}\makebox(0,0)[t]{\lineheight{1.25}\smash{\begin{tabular}[t]{c}Pinhole, $d$\end{tabular}}}}%
    \put(0.94463576,0.28304916){\color[rgb]{0,0,0}\makebox(0,0)[t]{\lineheight{1.25}\smash{\begin{tabular}[t]{c}Sample\end{tabular}}}}%
    \put(0.82624042,0.46100526){\color[rgb]{0,0,0}\makebox(0,0)[t]{\smash{\begin{tabular}[t]{c}Detector aperture\end{tabular}}}}%
    \put(0.83158186,0.3805408){\color[rgb]{0,0,0}\makebox(0,0)[t]{\smash{\begin{tabular}[t]{c}$\Omega$\end{tabular}}}}%
    \put(0,0){\includegraphics[width=\unitlength,page=2]{optical_system2.pdf}}%
    \put(0.78176338,0.24604771){\color[rgb]{0,0,0}\makebox(0,0)[t]{\lineheight{1.25}\smash{\begin{tabular}[t]{c}$f$\end{tabular}}}}%
    \put(0.51269059,0.00199994){\color[rgb]{0,0,0}\makebox(0,0)[t]{\lineheight{1.25}\smash{\begin{tabular}[t]{c}$L$\end{tabular}}}}%
    \put(0,0){\includegraphics[width=\unitlength,page=3]{optical_system2.pdf}}%
    \put(0.69442801,0.09935737){\color[rgb]{0,0,0}\makebox(0,0)[t]{\lineheight{1.25}\smash{\begin{tabular}[t]{c}$F$\end{tabular}}}}%
    \put(0.8521239,0.02809075){\color[rgb]{0,0,0}\makebox(0,0)[t]{\lineheight{1.25}\smash{\begin{tabular}[t]{c}$\phi$\end{tabular}}}}%
    \put(0,0){\includegraphics[width=\unitlength,page=4]{optical_system2.pdf}}%
  \end{picture}%
\endgroup%